# Structure and phase transitions into ionic adsorption layers on liquid interfaces

Roumen Tsekov
Department of Physical Chemistry, University of Sofia, 1164 Sofia, Bulgaria

The structure of ionic adsorption layers is studied via a proper thermodynamic treatment of the electrostatic and non-electrostatic interactions between the surfactant ions as well as of the effect of thermodynamic non-locality. The analysis is also applied to phase transitions into the ionic adsorption layer, which interfere further with the oscillatory-diffusive structure of the electric double layer and hydrodynamic stability of squeezing waves in thin liquid films.

Adsorption is an important phenomenon, which is responsible for many applications in food and cosmetic industry, flotation, etc. Usually, the adsorption of soluble surfactants on the air/water interface is described via the Langmuir-Blodgett monolayer concept originating from the two-dimensional physics of insoluble surfactants. The complexity of the Gibbs excess quantities leads to additional confusion.[1] The Gibbs adsorption is an integral over the concentration in the solution and thus the real distribution of the surfactant could be far away from the monolayer idealization. In the literature, there are many attempts to describe the adsorption of soluble surfactants as a result of specific attractive and repulsive forces between the dissolved molecules and the air/water surface.[2] They originate mainly from the interaction with the bulk water and the most important forces are dispersion, electrostatic and image ones. Since the dispersion forces decreases strongly by distance, the calculated adsorption layers are very thin, which corresponds well to the picture of insoluble surfactants. Recently, an excess interaction in thin liquid films was successfully attributed to the so-called adsorption disjoining pressure,[3,4] originating from the overlap of two adsorption layers. The latter could not be explained from the classical theory, where the adsorption layers are very thin. One of the aims of the present paper is to try to improve the classical adsorption models, which neglect the non-locality of the interactions between the surfactant molecules into the adsorption layer.

The theory of the electric double layer[5] dates back to the classical works of Helmholtz, Gouy, Chapman, Stern, Debye, Hückel and others. Due to general mathematical complications in charged systems, however, the applications are mainly restricted to dilute ionic solutions, where the electric potential is described via the Poisson-Boltzmann equation. The interactions between electric double layers are also intensively studied as an important component of inter-particle and colloidal forces.[6] Recently, significant attention has been paid to highly-charged Coulomb mixtures, where many specific phenomena take place[7] among them mono-species electric double layers.[8] General theories of charged fluids are also developed, which account for ion correlations going beyond the Poisson-Boltzmann theory.[9,10,11,12] An interesting aspect here is the effect of the non-electrostatic interactions between the ions in the electric double layer expected to become important in concentrated interfacial solutions.[13] Another aim of the present paper is to develop a simple theoretical approach to phase transitions in ionic adsorption layers based on electrostatics coupled with the Cahn-Hilliard model.[14]

To describe the adsorption of charged species in the frames of classical electrostatics one needs to calculate the local electric potential $\phi$, which satisfies the Poisson equation

$$\varepsilon_0 \varepsilon \partial_z^2 \phi = e(C - c) \tag{1}$$

Here $\varepsilon_0 \varepsilon$ is the dielectric permittivity of water, $C$ and $c$ are the local concentrations of monovalent surfactant anions and counterions, respectively. Usually, the solutes are considered to be spread according to the Boltzmann distribution, which is valid for dilute solutions only. Indeed, the concentration far away from the surface is low but near the surface it grows immensely driven by the surface forces. Hence, a good theory of soluble surfactants should necessary account for the intermolecular interactions into the adsorption layer. Moreover, an enormous gradient of the concentration appears near the surface as well, which indicates that the intermolecular interactions should be treated in a non-local manner. Since only the surfactant anions experience strong non-electrostatic attraction by the surface, the variation in the concentration of counterions is not dramatic and the surface is negatively charged. For this reason and to keep the consideration transparent, we will accept that the counterions concentration is nearly constant everywhere, equal to the bulk concentration. Such conditions can be experimentally reached by adding indifferent electrolyte. Thus, substituting in Eq. (1) the expression $c = C_\infty$, which follows from the electro-neutrality of the solution, yields a positive jelly model approximation

$$\partial_z^2 \phi = e(C - C_\infty) / \varepsilon_0 \varepsilon \tag{2}$$

At equilibrium the electrochemical potential of surfactant anions is constant everywhere. It can be generally expressed in the form

$$\tilde{\mu} = \mu_0 + k_B T \ln(\gamma C) + \lambda \partial_z^2 C + e\phi + \alpha / z^3 \tag{3}$$

where $\gamma$ is the activity coefficient. The Cahn-Hilliard third term[14] accounts for the non-locality of intermolecular interactions, while the last term in Eq. (3) describes the van der Waals attraction of the surfactant anions by the surface with $\alpha$ being proportional to the Hamaker constant. Due to the large hydration Born energy, there is a huge barrier at the surface preventing ions to penetrate into the gas phase. Without a detailed description we will account effectively for this repulsion between the surface and ions at small distances by a hard-core potential pre-venting the ions to get closer to the air/water surface than their hydration radii. Substituting the electric potential $\phi$ from Eq. (3) in Eq. (2) results in

$$\lambda \partial_z^4 C + k_B T \partial_z^2 \ln(\gamma C) + e^2 (C - C_\infty)/\varepsilon_0 \varepsilon + 12\alpha / z^5 = 0 \qquad (4)$$

Knowing the dependence of the activity coefficient $\gamma$ on the surfactant concentration closes the mathematical problem in Eq. (4). Our primary goal is not, however, to solve this strongly nonlinear equation, rather to discriminate between the different effects, affecting the structure of the adsorption layer. For instance, if the charge repulsion among the surfactant anions is the dominant force acting against the van der Waals attraction, Eq. (4) reduces simply to

$$C = C_\infty - 12\alpha \varepsilon_0 \varepsilon / e^2 z^5 \qquad (5)$$

Since $z < 0$ the surfactant concentration increases by approaching the surface inverse proportionally to the fifth power of the separation. Such a structure of the adsorption layer is expected to form at low temperature, where the effect of the entropy is negligible.

In the opposite case of negligible charge repulsion and especially in the case of non-ionic surfactants, Eq. (4) can be integrated twice to obtain

$$\lambda \partial_z^2 C + k_B T \ln(\gamma C / C_\infty) + \alpha / z^3 = 0 \qquad (6)$$

It is assumed here that $\gamma(C_\infty) = 1$ since the bulk solution is a dilute one. If one disregards the non-locality of the surfactant interactions ($\lambda \equiv 0$), Eq. (6) reduces to

$$\gamma(C) C = C_\infty \exp(-\alpha / z^3 k_B T) \qquad (7)$$

This equation describes a very sharp peak of $C$ near the surface, which practically coincides with the picture of insoluble surfactants. In the case of ideal solutions ($\gamma = 1$) Eq. (7) represents the Boltzmann distribution. In non-ideal solutions the peak is essentially reduced by the increase of the activity coefficient $\gamma$ but this effect cannot change dramatically the width of the adsorption layer. However, the existence of such a sharp concentration gradient near the interface emphasizes the importance of non-local thermodynamics. Thus, if the first term in Eq. (6) is the leading one, the concentration shows a very slow long-tail behavior

$$C = C_\infty - \alpha / 2\lambda z \qquad (8)$$

with a characteristic length $\alpha/2\lambda C_\infty$. Hence, the thermodynamic non-locality leads to spreading of the adsorption layer and, consequently, to the appearance of adsorption component of the disjoining pressure in thin liquid films.[4]

Let us pay now attention to other important phenomena, i.e. the phase transitions into the ionic adsorption layers. The Poisson equation (1) is nonlinear, which complicates essentially the electrostatic problem. However, it can be linearized to

$$\partial_x^2\phi + \partial_z^2\phi = \kappa^2\phi \tag{9}$$

where $\kappa$ is an effective inverse Debye length.[15] The latter reduces to the standard Debye parameter at low surface potentials. The solution of Eq. (9) is the following partial Fourier image

$$\phi_q = e\Gamma_\infty \theta_q \exp(Qz)/\varepsilon_0\varepsilon Q \tag{10}$$

where $\theta_q$ is the Fourier image of the local coverage by surfactant molecules of the surface and $\Gamma_\infty$ is their maximal adsorption. Since the electric potential does not diverge at minus infinity the real part of $Q \equiv \sqrt{q^2+\kappa^2}$ must be positive. The thermodynamic equilibrium into the interfacial layer requires constant value of the electrochemical potential $\tilde{\mu}_s$ at the surface, i.e.

$$\partial_x\tilde{\mu}_s = [k_BT/\overline{\theta}(1-\overline{\theta})-\beta]\partial_x\theta - \lambda_s\partial_x^3\theta + e\partial_x\phi_s = 0 \tag{11}$$

In this linearized expression $\overline{\theta}$ is the average coverage of the surface, $\beta$ is the Frumkin parameter accounting for the surfactant interactions and $\lambda_s \geq 0$ is the interfacial Cahn-Hilliard parameter, being proportional to the line tension between the dilute and condensed surface phases. Expressing from Eq. (10) the Fourier image $\phi_q(z=0) = e\Gamma_\infty\theta_q/\varepsilon_0\varepsilon Q$ of the surface potential $\phi_s$ and substituting it in the Fourier transformed Eq. (11) leads to the following equation

$$\lambda_s q^3 + [k_BT/\overline{\theta}(1-\overline{\theta})-\beta]q + e^2\Gamma_\infty q/\varepsilon_0\varepsilon Q = 0 \tag{12}$$

Since the coverage does not diverge at infinity, the imaginary part $q_{Im}$ must be zero, which leads automatically to $Q_{Re} > 0$ as well.

If the Frumkin parameter $\beta$ is negative or zero the surface phase is stable. In this case the solution of Eq. (12) reads $q=0$. Hence, the electrostatic potential in the bulk obeys the well-known exponential decay $\phi = e\Gamma_\infty\overline{\theta}\exp(\kappa z)/\varepsilon_0\varepsilon\kappa$, following from the Debye-Hückel approach. More interesting is when $\beta > 0$. In this case the effect of temperature becomes essen-tial. If the temperature is large enough $T > T_c \equiv \beta\bar{\theta}(1-\bar{\theta})/k_B$ the solution of Eq. (12) is $q=0$ again. This is not surprising since the surface phase is homogeneous and stable due to the entropy effect. However, if $T < T_c$ Eq. (12) possesses a non-trivial solution. Introducing two new characteristic constants $a \equiv (2\lambda_s \varepsilon_0 \varepsilon / e^2 \Gamma_\infty)^{1/3}$ and $A \equiv a^2[\lambda_s \kappa^2 + \beta(1-T/T_c)]/3\lambda_s$ one can rewrite Eq. (12) in the form

$$(aQ)^3 - 3A(aQ) + 2 = 0 \qquad (13)$$

Since $\lambda_s$ is proportional to $\beta/\Gamma_\infty$ the characteristic length $a \approx (2\beta\varepsilon_0\varepsilon / e^2 \Gamma_\infty^2)^{1/3}$ is typically of the order of a nanometer. The restrictions to Eq. (13) are $Q_{Re} \geq 0$ and $Q_{Im} \geq 0$. Hence, its proper solution reads

$$aQ = -(\sqrt{1-A^3}-1)^{1/3}(1+i\sqrt{3})/2 - [A/(\sqrt{1-A^3}-1)^{1/3}](1-i\sqrt{3})/2 \qquad (14)$$

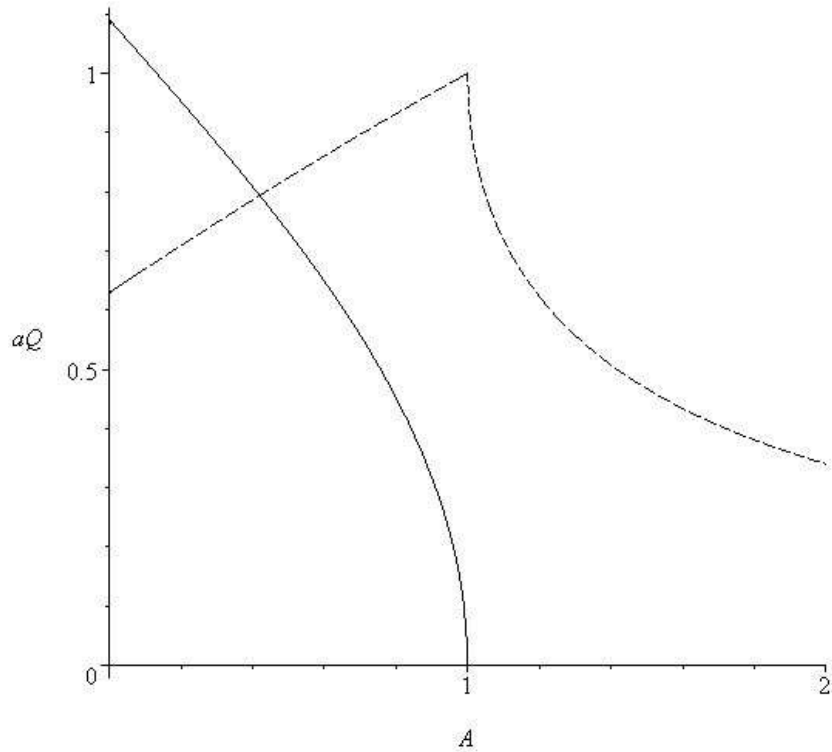

**Fig. 1** Wave vector's real (dash line) and imaginary (solid line) parts

The real and imaginary parts of the solution (14) are plotted in Fig. 1 as a function of the parameter $A$. Note that the latter increases with the decrease of temperature. As already mentioned at $T > T_c$ the wave vector components equal to $Q_{Re} = \kappa$ and $Q_{Im} = 0$. If tempera-ture falls below the critical one $T_c$, an interfacial phase transition appears on the surface. In the range $0 < A < 1$ the surfactant layer separates to dilute and condensed phases, which are distributed on the surface in a short-range order. The period of this structure is $Q_{Re}^{-1}$, while the decay length of the correlation is $Q_{Im}^{-1}$. Since the surface charge distribution is coupled with the bulk electrostatics, it appears that the electric potential in the bulk is oscillatory-decaying function with decay length $Q_{Re}^{-1}$ and oscillation length $Q_{Im}^{-1}$. If the temperature falls further a liquid crystal structure covers completely the surface at $A > 1$. It is a perfect crystal with $Q_{Im} = 0$ and lattice constant $Q_{Re}^{-1}$. The latter is also the decay length of the electric potential in the bulk, which is no more oscillating. Note that in this case the decay length $Q_{Re}^{-1} \sim a$ is much shorter than the Debye one $\kappa^{-1}$. An interesting feature of the transition between the crystal-like and liquid-like structures at $A = 1$ is that the temperature coefficients of the characteristic lengths show discontinuity (see Fig. 1), thus indicating a kind of a second order phase transition.

As was mentioned, thin liquid films are important colloidal systems, affected by the structure of the ionic adsorption layer. Since foam films are symmetric, the electrostatic potential $\phi(z) = \phi(-z)$ is also symmetric. The linearized solution of Eq. (9) for films reads

$$\phi/\phi_s = \frac{\cosh(\kappa z)}{\cosh(\kappa h/2)} - \frac{\kappa \tanh(\kappa h/2)}{2} \sum_q \frac{\cosh(Qz)}{\cosh(Qh/2)} \zeta_q \exp(iqx) \qquad (15)$$

where $\phi_s$ is the surface potential, $h$ is the film thickness and $\zeta_q$ is the Fourier image of the small amplitude of the surface waves. The overlap of the two electric layers results in appearance of electrostatic disjoining pressure. It can be calculated from the electric potential $\phi_0$ in the middle of the film via the expression $\Pi_{EL} = \varepsilon_0 \varepsilon \kappa^2 \phi_0^2 / 2$. Introducing here the electric potential from Eq. (15) yields

$$\Pi_{EL} = \frac{\varepsilon_0 \varepsilon \kappa^2 \phi_s^2}{2\cosh^2(\kappa h/2)} - \frac{\varepsilon_0 \varepsilon \kappa^3 \phi_s^2 \sinh(\kappa h/2)}{2\cosh^3(\kappa h/2)} \sum_q \frac{\cosh(\kappa h/2)}{\cosh(Qh/2)} \zeta_q \exp(iqx) \qquad (16)$$

The first term in Eq. (16) is the electrostatic component of the disjoining pressure in the case of flat film surfaces, i.e. $\bar{\Pi}_{EL} \equiv \varepsilon_0 \varepsilon \kappa^2 \phi_s^2 / 2\cosh^2(\kappa h/2)$. One can easily check that the multiplier of the second term is equal to the derivative $\bar{\Pi}'_{EL} \equiv \partial_h \bar{\Pi}_{EL}$. Thus, Eq. (16) acquires the form

$$\Pi_{EL} = \bar{\Pi}_{EL} + \bar{\Pi}'_{EL} \sum_q \frac{\cosh(\kappa h/2)}{\cosh(Qh/2)} \zeta_q \exp(iqx) \qquad (17)$$

In accordance to Eq. (17) the dispersion relation of the squeezing modes of the hydrodynamic corrugations in the film acquires the form[16]

$$24\eta i\omega / h(qh)^2 = \sigma q^2 - \bar{\Pi}'_{VW} K_2(qh)(qh)^2 - 2\bar{\Pi}'_{EL} \cosh(\kappa h/2)/\cosh(Qh/2) - 2\bar{\Pi}'_{AD} \quad (18)$$

Here $\omega$ is the frequency of the waves, $\eta$ is the liquid viscosity, $\sigma$ is the surface tension on the air/water interface, $\bar{\Pi}'_{VW} = 3K_{VW}/h^4$ is the thickness derivative of the van der Waals component of the disjoining pressure and $K_2(\cdot)$ is the modified Bessel function of second kind and second order. The novel part in Eq. (18) is that the thickness derivative of the electrostatic component of the disjoining pressure is multiplied by the factor $\cosh(\kappa h/2)/\cosh(Qh/2)$. Thus, the already discusses phase transitions into the ionic adsorption layer will affect via $Q$ also the dispersion relation of the surface waves and the stability of the thin liquid films. The new heuristically added adsorption component of the disjoining pressure can be estimated, for instance, from Eq. (9) to obtain $\bar{\Pi}'_{AD} = -k_B T\alpha/\lambda h^2$.